\documentclass[12pt]{article}
\usepackage{graphicx}
\usepackage{enumerate}
\usepackage{url} 
\usepackage[utf8]{inputenc}
\usepackage{float}
\usepackage{amsmath}
\usepackage[round]{natbib}
\usepackage{hyperref} 
\usepackage{lscape}
\usepackage{bm}
\usepackage{nameref}
\usepackage{setspace}
\usepackage{titlesec} 
\usepackage{xcolor}
\usepackage{subfig}
\usepackage{authblk}
\usepackage{ulem}
\usepackage{tabularx}

\definecolor{orange}{RGB}{255,140,0}

\definecolor{blueberry}{RGB}{0,0,153}

\title{Bridging the Gap Between Methodological Research and Statistical Practice: Toward "Translational Simulation Research}
\author{Anne-Laure Boulesteix$^{1,2*}$, Patrick Callahan$^{1}$, Luzia Han\ss um$^1$, \\Vincent Gärtner$^3$,
Eva Hoster$^1$\\ }

\begin{document}

\maketitle

\noindent
$^1$: Institute for Medical Information Processing, Biometry and Epidemiology, Faculty of Medicine, LMU Munich, Germany\hfill\break\\
$^2$: Munich Center for Machine Learning, Munich, Germany\hfill\break\\
$^3$: Department of Neonatology, Dr von Hauner University Children's Hospital, LMU Munich, Germany\hfill\break \\
$^*$: corresponding author, boulesteix@ibe.med.uni-muenchen.de
\hfill\break\\


\newpage

\begin{abstract}
\noindent
{\bf Background:} 
Simulations are valuable tools for empirically evaluating the properties of statistical methods and are primarily employed in methodological research to draw general conclusions about new or existing approaches. In addition, they can often be useful to applied statisticians and data analysts, who may rely on published simulation results to select an appropriate statistical method for their application. However, on the one hand, applying published simulation results directly to practical settings is, in our experience, frequently challenging, as the scenarios considered in methodological studies rarely align closely enough with the characteristics of specific real-world applications to be truly informative.
 Applied statisticians, on the other hand, may struggle to construct their own simulations or to adapt methodological research to better reflect their specific data  due to time constraints and limited programming expertise. \\ {\bf Methods:} We propose bridging this gap between methodological research and statistical practice through a translational approach by developing dedicated software along with simulation studies, which should abstract away the coding-intensive aspects of running simulations while still offering sufficient flexibility in parameter selection to meet the needs of applied statisticians. \\ {\bf Results:}
We demonstrate this approach using two practical examples, illustrating that the concept of translational simulation can be implemented in practice in different ways. In the first example---simulation-based evaluation of power in two-arm randomized clinical trials with an ordinal endpoint---the solution we discuss is an R package called {\tt ordinalsimr}, which includes a Shiny web application providing a graphical user interface for running informative simulations in this context. For the second example---assessing the impact of measurement error in multivariable regression---a streamlined, less labor-intensive approach is suggested, involving the provision of user-friendly, well-structured, and modular analysis code.\\
{\bf Conclusion:} Regardless of the specific form its implementation takes, we believe that when methodological researchers consciously incorporate the concept of translational simulation into their work, it can play a key role in bridging the gap between methodological research and statistical practice.
\end{abstract}

\newpage
\section{Background}

\label{sec:intro}
Simulations are useful for empirically assessing the properties of statistical methods \citep{burton2006design,morris2019using,boulesteix2020introduction}. They are mainly used in methodological research to draw conclusions about the performance of new or existing methods in different settings. Rather than aiming to identify a universal winner method that \lq\lq beats them all'', simulations are better defined as attempts to query in {\it which settings which methods} work better, as argued by \cite{strobl2024against} who warn against the \lq\lq one method fits all datasets philosophy''. \cite{friedrich2024role} contrast simulation studies against empirical evaluations based on real data, also termed \lq\lq benchmarking'', and argue in favor of combining the advantages of both approaches. 
Such works on simulation methodology take the perspective of {\it methodological researchers} whose aim is to gain new insights on the methods behavior for a {\it class} of applications. Regardless of whether the targeted application domain is narrow or broad, methodological researchers seek to gain insights that extend beyond a specific dataset or study.

Less discussed is the role of simulations for applied statisticians who plan and conduct data analyses in fields of application of statistics, such as medicine, from which our examples are drawn. Such applied statisticians are interested in a particular (medical) research question and study, but {\it not} in the performance of methods for other studies or for other research questions.  They may  employ simulations tailored to their specific application to objectively compare methods based on properties such as estimator bias or test power \citep{boulesteix2020introduction}, and to guide the development of an analysis plan \citep{nance2024causal}. For example, it may be meaningful to assess the behavior of a method under particular conditions that deviate from the standard situation and are (likely) relevant to the considered application. See \cite{brakenhoff2018random} for a  study on the impact of measurement error in multivariable regression and the more complex study by \cite{abrahamowicz2024data} on the impact of data imperfections such as missing confounder data and imprecise timing of an interval-censored event in survival analysis. 

Even though such simulation studies may be extremely useful, they are rarely conducted in practice in the context of medical research projects for a variety of reasons. One contributing factor may be that the medical research community---including journal editors, reviewers, and readers—--often perceives them as overly complex (particularly when explanations are limited by word count restrictions) and does not understand their pivotal role in study planning. We conjecture, however, that the lack of time resources and/or the lack of deep expertise on simulation methodology from the side of many study statisticians play a major role. This may lead to the use of suboptimal methods that fail to take the complexity of the setting into account and ultimately yield biased or less informative results.  For example, the simulations conducted by \cite{abrahamowicz2024data} in the context of their two showcases yield valuable insights that may---if taken into account throught the choice of an appropriate method---considerably improve the reliability of the study results.

Ideally, simulation results from methodological research could assist applied statisticians who, e.g. due to time or resource limitations, are unable to conduct tailored simulations for their own data settings. In our experience, however, this rarely occurs in practice. One factor may be that the reporting of statistical simulation studies  often lacks relevant information \citep{kelter2024bayesian}. Another important problem is that the settings considered in methodological publications rarely fit the characteristics of applied scenarios closely enough to be of direct utility. In some cases, the simulation settings used in methodological studies may be overly simplistic or of limited focus; even when they are not, the number of reported settings is inevitably limited. All in all, even when methodological researchers publish high-quality simulation studies, applied researchers will rarely find simulation results that answer their particular questions about method choice in the specific study they have to plan. In other words, translating simulation results from methodological literature into statistical practice poses a significant challenge.

To fill this gap, we propose the concept of \lq\lq translational simulation research'' (in analogy to \lq\lq translational medical research'') and demonstrate its implementation through two simple examples. These showcases demonstrate how methodological researchers reporting simulation studies can provide practical help to an applied statistician or data analyst who 1) does not find simulation methodological results that would adequately inform their choice of  analysis method, and 2) is not able or has no time to conduct a simulation study by themselves. 

As a first showcase, we consider the following example. At the planning stage, an applied statistician has to select an analysis method for a two-arm randomized clinical trial (RCT) whose primary endpoint is a variable measured at an ordinal scale. Methodological literature on the planning and analysis of such trials is relatively scarce \citep[see][for a recent overview]{selman2024statistical}. It is thus likely that our statistician will not find any simulation results in the methodological literature giving the power of the potential statistical tests for the specific distributions of the ordinal endpoint (in the control and intervention arms) that are relevant to their study. The Shiny app we introduce in this paper fills this gap, and provides guidance for the choice of a statistical test with appropriate power based on the assumptions made by its users. 

As a second showcase, we consider a simulation addressing the impact of measurement errors in the context of regression analysis of observational studies as studied by \cite{brakenhoff2018random,boulesteix2020introduction}. It illustrates a simpler, less labor intensive approach, which consists of providing user-friendly and well-documented analysis codes especially designed to allow applied statisticians to easily re-run the simulation with their own parameter settings without requiring advanced programming skills. 

The remainder of this paper is structured as follows. The Methods section outlines in detail the concept of translational simulation research. The Results section is devoted to the examples, while the Discussion section addresses connections to other related ideas and limitations of the concept of translational simulation research.

\section{Methods}
\subsection{Translational research in other fields}
The term \lq\lq translational research'' may have different meanings even within a single field, such as medicine. For example, for many researchers from the medical field, the term refers to \lq\lq the \lq\lq bench-to-bedside'' enterprise of harnessing
knowledge from basic sciences to produce new drugs, devices, and treatment options for patients''  \citep{woolf2008meaning}.  Alternatively, some understand it as the effort to guarantee that research results (e.g., new drugs) lead to real-world patient benefits. Regardless of the exact definition, there is a general consensus that translation of knowledge from basic sciences into practice is crucial to scientific and technological progress.

\subsection{The gap between methodological research and statistical applications}
When considering  methodological simulation-based evaluations of statistical methods as a basic science, and applications to (medical) data as its \lq\lq bedside'' counterpart, it becomes obvious that
\begin{itemize}
    \item this basic science has huge translational potential because most results on the performance and behavior of statistical methods are potentially relevant for a class of applications;
    \item this potential is far from being fully exploited: we conjecture that many simulation results are read by methodological researchers rather than used by applied statisticians to plan specific studies, as suggested by citation records of methodological articles.
\end{itemize}
This disconnect often results in frustration on both sides \citep{sauerbrei2014strengthening}. Methodological researchers  often feel their research has poor impact, and applied researchers often feel methodological literature does not provide them the information they need. 
This problem is certainly related to the fact, following the terminology of \cite{heinze2024phases}, methodological literature tends to focus too much on early phases of methodological development. Following \cite{heinze2024phases}, these phases {\it \lq\lq cover (I) proposing a new methodological idea while providing, for example, logical reasoning or proofs, (II) providing empirical evidence, first in a narrow target setting, then (III) in an extended range of settings and for various outcomes, accompanied by appropriate application examples, and (IV) investigations that establish a method as sufficiently well-understood to know when it is preferred over others and when it is not''}. 

In a perfect world, we would have many detailed phase IV studies that could be used by applied statisticians to plan their studies. However, a world in which sufficient available evidence for all imaginable settings is more of a utopia than a realistic scenario. Referring to the example of ordinal endpoints we will present in Section~\ref{sec:example1}, it is unrealistic that simulation results are available for any possible distribution of the ordinal endpoint in the two treatment arms. This would require (i) a huge computational expense (which does not make sense since most of these results will never be used by anybody), (ii)  infrastructure efforts to make these results available to applied statisticians in a user-friendly way, (iii) and a major change of incentive structure as such extensive simulation studies on existing methods are not  currently (sufficiently) recognized as valuable scientific research \citep{heinze2024phases,boulesteix2017towards}. While there may be positive developments in the future regarding points (ii) and (iii), (i) will always remain a problem in a world with limited resources.

\subsection{Translational simulation research: definition and implementation}
The concept of translational simulations fits in this context. A translational simulation is a simulation implemented by methodological researchers in such a way that its use by applied statisticians planning their studies is facilitated. More precisely, for a simulation study to qualify as translational, its readers should be able to easily and rapidly re-run the simulation for the specific setting that is relevant to the application study they are working on.

From a technical point of view, this goal can be achieved in different ways. Interactive web tools (e.g., Shiny apps) are particularly appealing for the target audience of data analysts with limited programming expertise. We will briefly present such an example in the Results section (example 1). Another solution consists of developing code for simulations using modular components (data generation, analysis, performance measurement) and then tying them all together into a \lq\lq simulation driver'' function that can execute the entire simulation for one condition, as generally recommended by \cite{miratrix2024designing}. To support the setup of such a modular simulation in complex contexts, dedicated packages have been developed, such as {\tt simhelpers} \citep{simhelpers} or {\tt SimDesign} \citep{SimDesign}.
Such a modular framework can be implemented along with literate programming tools such as Quarto, which make it easy to dynamically generate reports for additional simulation settings.

The concept of translational simulation, however, is not inherently tied to Shiny apps, literate programming, or specific modular frameworks. A minimalistic implementation that avoids such tools is presented in the Results section (example 2), where the simulation code is simply divided into two files: one for user-friendly specification of simulation parameters, and another for executing the simulation based on those parameters.

\section{Results}
\subsection{Example 1}
\label{sec:example1}

\subsubsection{Addressed methodological question}
In recent years, the value of ordinal endpoints—compared to commonly used continuous or binary endpoints—has attracted growing interest among medical researchers conducting clinical trials \citep{d2020ordinal,ceyisakar2021ordinal,selman2024statistical}. In particular, ordinal endpoints do not lead to the same loss of information as dichotomized endpoints and offer major advantages in terms of power \citep{ceyisakar2021ordinal,selman2024statistical}. 
\cite{selman2024statistical} report that the proportional odds model—supported by simple closed-form formulas for power and sample size \citep{whitehead1993sample, kieser2020methods}—is commonly used in practice to analyze ordinal endpoints. However, some authors instead employ the Mann–Whitney U-test, the chi-square test, or Fisher’s exact test, while others dichotomize their ordinal endpoints \citep{selman2024statistical}.

While our goal is not to provide an extensive literature review in the context of this paper devoted to the concept of translational simulations, we can confidently say that (i) statistical planning
and analysis of such endpoints is not yet widely established and practical guidance is scarcer for ordinal endpoints than for continuous, binary or survival endpoints,  and (ii) simulation studies presented in existing methodological literature can rarely inform precisely the choice of a statistical method in a specific application because the (mostly simple) distributions of the ordinal endpoint in the trial arms often do not reflect the reality of the considered trial.

\subsubsection{Example study}
\label{subsec:example1}
\lq\lq Desirability of outcome ranking'' (DOOR) is a class of ordinal endpoints that has gained increased attention from the clinical trial world in recent years \citep{evans2015desirability}. For example, in the context of infections treated with antibiotics, DOORs take both the clinical outcome and the duration of treatment into account \citep{evans2015desirability} in a hierarchical way. The underlying idea is that, the outcome being equal, a shorter treatment duration is desirable. 
The methodology has also recently shown potential in the field of neonatology, where it is claimed to yield a more meaningful assessment of the benefits and harms of novel interventions with greater insight for clinicians, infants and parents than typical composite measures \citep{katheria2024application}.

As an example, we take a two-arm randomized trial in neonatology considering a DOOR endpoint with six categories as defined in the second column of Table~\ref{tab:gaertner}. Of note, these categories were chosen as being possibly related to an early respiratory intervention (i.e. changes in respiratory management during stabilization of preterm infants in the delivery room and/or the first hours after birth). For different neonatal interventions, other major morbidities may need to be considered. Also, the final outcome categories were ranked in an iterative process in conjunction with parents and patient representatives from the \lq\lq Bundesverband das Frühgeborene Kind e.V.''. 

We take the perspective of an applied statistician planning a randomized controlled trial with this DOOR endpoint. Obviously, the choice of a statistical test  depends on what the clinical experts expect in both treatment arms. Based on previous related studies targeting similar populations and their own experience, neonatologists assume that in the control arm the DOOR endpoint will approximately follow the probability distribution specified in the column \lq\lq Control''  of Table~\ref{tab:gaertner}. Column \lq\lq Intervention'' contains the corresponding expected probabilities in the intervention group. 

This probability distribution neither clearly fulfills the distributional assumptions of the Mann-Whitney U-test (same distribution shape in the two arms) nor those of the proportional odds model, making the choice of the statistical test difficult. In this context, it may make sense to rely on simulations to conduct this choice, but there is obviously no methodological study in the literature that the trial statistician might refer to and that considers precisely the probability distributions from Table~\ref{tab:gaertner}. In the next section, we describe a translational approach that methodological authors could  adopt when reporting simulation studies on the power and type 1 error of tests for two-arm trials with ordinal endpoints, to make their studies more useful to applied statisticians.


\begin{table*}
\centering
    \begin{tabular}{llcc}\hline
        Rank &  Description & Control & Intervention  \\ \hline
               Rank 1 & Alive and well, i.e. none of the below mentioned morbidities & 26.5\% & 47.5\%  \\
       Rank 2 & Alive without major or moderate morbidities but with mild& 27.5\% & 18.0\% \\  &  morbidity: Respiratory failure within the   first 72 hours  &  &  \\
        Rank 3 & Alive without any of the two major morbidities (see Rank 4)  & 24.7\% & 15.0\%  \\ & but with any of the following moderate  morbidities: & & \\ & Mechanical ventilation $>$ 72 hours overall or IVH I/II° & & \\ & or pneumothorax requiring drainage & & \\  
          Rank 4 & Alive with one of the two major morbidities: Moderate/ & 15.1\% & 13.7\%  \\ & severe bronchopulmonary dysplasia or severe brain lesion  &  & \\
       Rank 5 & Alive with the two major morbidities (see Rank 4) 
 &  2.0\% & 1.8\% \\
Rank 6 & Dead &  4.2\% & 4.0\% \\
        \hline
    \end{tabular}
    \caption{Definition of the 6 possible DOOR categories (column 2), assumptions for their probability in the control group (column 3) and intervention group (column 4).}
    \label{tab:gaertner}
\end{table*}


\subsubsection{Proposed solution: a Shiny application}
\label{subsec:app}

\begin{figure*}
    \includegraphics[width=18cm]{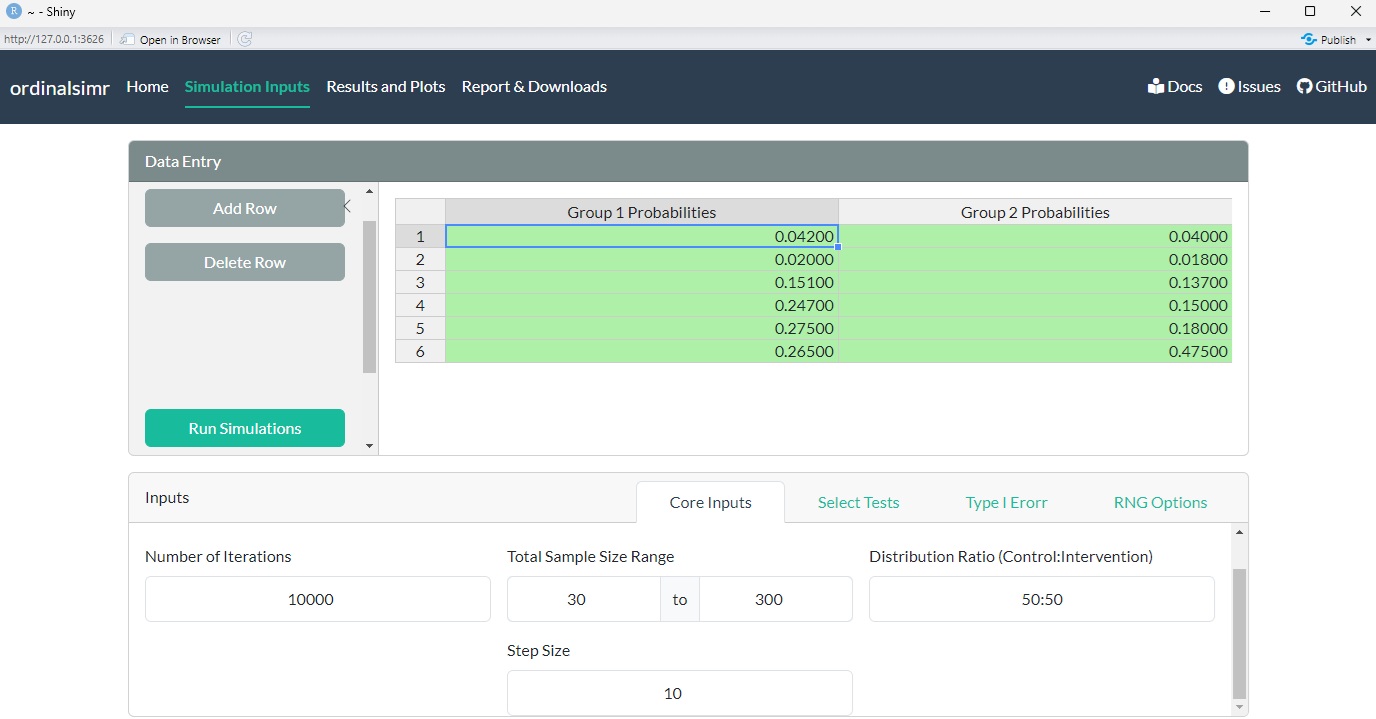}
    \caption{User interface of the Shiny app ordinalsimr: simulation inputs.}
    \label{fig:enter-label}
\end{figure*}

Authors presenting comparison studies of tests for ordinal endpoints may implement a Shiny app allowing applied statisticians to repeat the simulation with their own settings.
Shiny \citep{shiny} is a free and open-source framework that enables data analysts to easily  build interactive web applications straight from R and, more recently, from Python. 
To demonstrate the use of such applications towards translating simulation research into practice, one of us built the  R package {\tt ordinalsimr} \citep{callahan_ordinalsimr_2025}, which is publicly available from CRAN for R versions $>=4.4.0$ and whose source code is available on GitHub at \url{https://github.com/NeuroShepherd/ordinalsimr} along with a detailed documentation. A limited memory live app is hosted for demoing purposes at: \url{https://6fcd1k-pat.shinyapps.io/ordinalsimr/} for demoing purposes.
The app is designed to support trial statisticians planning a two-arm RCT with an ordinal endpoint in conducting simulation studies that evaluate the type I error and power of tests potentially applicable in this context and already available in R \citep[see][for more details]{callahan_ordinalsimr_2025}. Implemented as a Shiny application, the package abstracts away the more complex coding typically required to set up such simulation studies, offering an accessible interface for specifying key parameters. Users can input directly into the interface the key parameters for which they want to do power analysis, including the considered testing methods, the set of considered sample sizes, the allocation ratio between control and intervention arms, the expected distribution of the ordinal endpoint  for the control arm (column \lq\lq Control'' in the example from Table~\ref{tab:gaertner}) and the intervention arm (column \lq\lq Intervention''), and the desired number of simulation repetitions (typically, $\geq$ 10,000 to achieve satisfactory precision in type 1 error and power approximation). The user interface is diplayed in Figure~\ref{fig:enter-label}. The results (approximate type-1 error and power for the considered methods and sample sizes) are displayed in tabular and graphical format and saved as an {\tt .rds} file.

\subsection{Example 2}
\label{sec:example2}

\subsubsection{Addressed methodological question}
 In epidemiological studies, the relation between an exposure and an outcome of interest is often investigated using regression analysis. Measurement error can affect the exposure, the outcome as well as confounding variables and potentially impacts the estimation of the effect of the exposure on the outcome. In epidemiological literature, measurement error is often claimed to lead to an underestimation (i.e. attenuation) of this effect \citep{brakenhoff2018measurement}, aligning with the intuition that noise obscures the underlying signal of interest.
 However, the situation becomes more complex when measurement error affects at least one confounding variable.  The magnitude and direction of the bias depends on the joint distribution of the considered variables and on the type and severity of measurement error.
 As demonstrated by \cite{brakenhoff2018measurement,boulesteix2020introduction}, measurement error can also lead to an overestimation of the effect of interest.
 
\cite{boulesteix2020introduction} consider an exemplary hypothetical observational study of the association between glycated haemoglobin (HbA1c) levels and systolic blood pressure (SBP) based on a publicly available dataset from the National Health and Nutrition Examination Survey (NHANES 2015-2016). Multiple linear regression is used to investigate the effect of HbA1c (considered as exposure) on SBP (considered as outcome) while adjusting for the confounders age, gender and body mass index (BMI). The positive estimated effect of HbA1c on SBP is lower after adjustment for BMI than without adjustment. 

\cite{boulesteix2020introduction} show a small simulation study that assesses the impact of measurement error on the estimation of the BMI-adjusted effect in this context. More precisely, for the considered real data set they artificially add random measurement error to the exposure HbA1c, to the confounder BMI or to both and re-estimate the effect of interest for the resulting dataset. This procedure is repeated numerous times across varying levels of measurement error. The results can be summarized as follows. While measurement error in the exposure HbA1c leads to an attenuation of the estimated effect, measurement errors in the confounder BMI leads to overestimation. Further details, along with the code used to implement the simulation, are provided in the supplementary materials of \cite{boulesteix2020introduction}. 

\subsubsection{Example study}
Imagine a scientist conducting an observational study investigating the effect of an exposure \lq\lq meat consumption'' on an outcome reflecting cardio-vascular health while adjusting for confounders such as age, BMI, or cigarettes per day. They have already collected/extracted the data set to be analysed and are interested to assess the impact of measurement error in their setting to better interpret their results. In principle, this can be achieved by conducting a simulation study similar to the ones by \cite{brakenhoff2018random} and \cite{boulesteix2020introduction}. However, the simulation results from these studies are not directly applicable to our scientist’s context, as the joint distribution of exposure, outcome, and confounders in their application differs from those considered in the original simulations by \cite{brakenhoff2018random} and \cite{boulesteix2020introduction}. In the next section, we explain what \cite{boulesteix2020introduction} could have done to support the researchers conducting the study on the effect of meat consumption on cardiovascular health.

\subsubsection{Proposed solution: appropriately documented and structured modular simulation code}
The solution we demonstrate here consists of providing code specifically designed to be accessible to applied statisticians without deep expertise in simulation studies. By \lq\lq accessible'', we mean that the code should clearly guide users to the sections where they need to input their data and/or application-specific simulation parameters, allowing them to identify these points at a glance. This solution is less labor intensive for the methodological research than the app presented in Example~1.

The supplement of this paper includes two examples files illustrating this approach, example2$\_$core.R and example2$\_$input.R. The user-friendly file example2$\_$input.R file is intended to be completed and run by users, allowing them to specify the data and parameters relevant to their application-specific simulation. The documentation of the code guides them through this process. For example, in the case study on the impact of measurement error described above, the user must import their dataset into R, format it according to the provided instructions (e.g., standardizing variable names as described in the documentation and ensuring the data is in a data frame format), and define the values to be investigated for the magnitude of measurement error in the exposure and confounders—expressed as a proportion of variance. Once this step is completed, they can run example2$\_$core.R, that automatically loads example2$\_$input.R in order to run the simulation with the parameters specified by the user.



\section{Discussion}
\subsection{Related but different concepts}

\subsubsection{Reproducibility}
The importance of research reproducibility has  been increasingly emphasized over the past decades, including in methodological research. See for example the reproducibility policy of the Biometrical Journal \citep{hofner2016reproducible}. We share the view that code enabling the reproduction of all simulation results with a single mouse click should be made available alongside any article presenting a simulation. 

The concept of translation simulation as implemented in Example~2 is related to the concept of reproducibility. If readers have access to files such as example2$\_$core.R and example2$\_$input.R, they may use them to reproduce results of the entire simulation study, by entering all considered parameter settings successively in the input file. However, this would be a very inefficient way to reproduce the entire study! Conversely, if the authors make high-quality and well documented code available for the purpose of reproducibility, it is possible, at least for an expert programmer, to adapt the code in such a way that it runs for a few additional application-specific settings. 

However, the distinct objectives of translational simulation and reproducibility demand fundamentally different implementations, both of which should ideally be provided by methodological researchers when publishing a simulation study. To ensure reproducibility, the code must be designed so that the study’s complete results can be regenerated with a single mouse click. Ideally, the simulation should be efficiently programmed to minimize total computing time—which can be considerable, given that it encompasses a multitude of settings.  In contrast, translational simulations prioritize user-friendliness over computational efficiency, which is less critical in this context since users typically run the simulation for only a limited number of settings. The user is responsible for specifying these settings, and the code should be designed to make this process as straightforward as possible.

\subsubsection{Interactive visualization of results}
An approach that is becoming increasingly common \citep[see for example][]{hanke2024variable}, consists of making extensive simulation results \citep[or real data benchmarking results, ][]{probst2019tunability} available through an interactive web app. In contrast to our concept of translational simulations,  the results have {\it already} been produced by the researchers who conducted the methodological study. The app simply allows users to call and visualize them in a user-friendly manner rather than to produce them. It is an elegant and practical alternative to the tedious presentation of large confusing tables and to good-looking but imprecise and unwieldy plots. 


\subsubsection{Platforms for neutral comparisons}
Interactive platforms facilitating the systematic neutral comparison of statistical methods are another promising class of tools. Such tools, which exist in different variants and levels of complexity, are primarily intended for {\it methodological} researchers. The idea here is to provide access to a simulation infrastructure (in a broad sense) to researchers who are interested in objectively comparing a (new) method to existing ones. This may include meaningful simulation designs, high-standard implementations of existing competing methods and/or the computation of various performance measures. Examples of such platforms have been recently described as part of the special collection entitled \lq\lq  Towards neutral comparison studies in methodological research'' published in Biometrical Journal \citep{kodalci2024neutralise,ruberg2024platform}. 

These platforms are also fundamentally different from our proposal in that they target a methodological audience and aim at an evaluation of methods for a class of applications across real data sets and/or simulation settings. For example, the principle of the open science initiative by \cite{kodalci2024neutralise} dedicated to simple two-sample tests (such as the t-test) is that \lq\lq everyone can submit a new method and/or a new simulation scenario, and the system evaluates (1) the new method on all previously submitted simulation scenarios, and (2) all previously submitted methods on the new scenarios.''  
In the same vein, \cite{ruberg2024platform} create a platform for the evaluation of methods aiming at identifying subgroups from clinical trial data. In particular, this platform includes a common data generating mechanism for appropriately simulating realistic data as well as a common evaluation system for comparing the methods' performances. Both \cite{kodalci2024neutralise} and \cite{ruberg2024platform} aim at making {\it methodological} comparison studies more reliable and fair rather than primarily filling the needs of applied statisticians working on a specific application---even if high quality methodological research indirectly also serves the purposes of applied statisticians in the long run. 

\subsubsection{Software for simulating realistic data}
There are numerous packages implementing the simulation of complex data of various types with the aim to enable their users to easily compare the performance of methods. While many of these packages target methodological researchers who are interested in a whole domain of applications and will typically vary the data parameters over a broad range, some rather have applied statisticians in mind; see   \cite{wahab2024cities} for a recent example. Their Shiny application  called 'CITIES' simulates data that \lq\lq resemble real-life clinical trials with respect to their reported summary statistics, without requiring the use of the original trial data'' and allows method comparisons yielding causal inference in the presence of intercurrent events. In the same vein, \cite{pustejovsky2014alternating} introduce an R package that allows simulation of behavior streams and data from several recording procedures in the context of behavioral sciences.
Such applications and packages can be used as part of the implementation of the translational simulation studies we advocate in the present paper.

\subsection{Limitations}
An obstacle to the widespread adoption of translation strategies in simulation studies by methodological researchers is the time and effort required. We acknowledge this limitation, but also note that including a translation strategy may increase an article’s visibility and citations, providing both incentive and recognition for authors. Furthermore, large language models might help reduce the time and effort required to implement such strategies.

One may also argue that there is no need for translational simulations provided those who plan studies receive appropriate education allowing them to implement simulations autonomously. We agree that it generally makes sense to provide training on simulations to analysts who do not do themselves methodological research \citep{boulesteix2020introduction}. Especially if simulations  presented in the literature are accompanied with open code for the purpose of reproducibility, it may in some cases be possible even for analysts without strong methodological background to re-run the simulation in their own setting. However, we claim that, even in this better world where all data analysts have increased simulation expertise and skills and all simulations are reproducible, translational simulations may still be a valuable concept. In our experience, it is unlikely (and not desired) that  trial statisticians spend days to understand the code details of the simulation of interest. Moreover, without a deep understanding of the code, the process of re-running the simulation is prone to errors. That is because simulation code is most often not developed with the goal to reduce the probability of errors when run by laypersons, but rather to achieve efficiency in terms of computation time and storage. This is particularly true when the code is parallelized and optimized using complex coding techniques to increase computational efficiency.  Any error is unacceptable in the context of the planning of a medical study. That is why we claim that it is necessary to build bridges between simulations performed within methodological research with complex code on one hand, and applied statisticians who have another spectrum of competences and limited time resources on the other hand.

One may also argue that we would not need translational simulations if there were enough accumulated evidence allowing to derive clear-cut guidance without running new simulations. Ideally, this guidance might take the form of a decision tree algorithm suggesting the applied statistician a recommended method based on his answer to a series of question. Regardless whether such advance is realistic or not, our position is as follows.
For such reliable evidence-based guidance to make our concept obsolete, it should have a very high resolution. By this, we mean that many different parameter settings are considered so that it is reasonable to interpolate results in places of the parameter space that were not covered by simulations conducted so far. This, in turn, implies a huge computational effort, which will be in large part useless, because many parts of the parameter space will never be relevant to any study. 
Remember the DOOR endpoint in neonatology trials, which takes six possible values. Taking into account that the probabilities of the six values sum to 1, we have five degrees of freedom when specifying the probability distribution of the endpoint in the control arm, and also five degrees of freedom for the probability distribution of the endpoint in the intervention arm. Developing a decision algorithm that remains valid across the entire ten-dimensional parameter space appears unrealistic---or at the very least, extremely difficult and currently unfeasible. 

Finally, an important criticism toward our concept is that it is barely feasible to build very simple apps with an intuitive user interface for implementing complex simulations. We admit that our concept is not appropriate for all types of simulation studies, especially not for the most complex ones. If the simulation design is very complex, understanding it will require a level of expertise that can be only reached by statisticians who are also able to re-run the simulation themselves provided code is available. Moreover, in many cases the specific scenario relevant to the applied statistician will require changes in the simulation code, which again can only be performed by an expert of statistical simulations. In such cases, the efforts of methodological researchers are likely better spent on improving code documentation to enhance reproducibility, rather than on developing an app. Similarly, investing time and resources may not be justified during the early stages of methodological development \citep[phases I and II in the terminology by][]{heinze2024phases}---particularly when simulations are intended only to provide a preliminary proof of concept for a new method. Exactly as no statistical method is expected to beat all other methods for all data contexts \citep{strobl2024against}, we do not claim that our concept of translational simulation is not the panacea in all simulation contexts. But it is certainly valuable in many of them.

\section{Conclusion}
Simulation studies can often inform applied statisticians for the choice of appropriate statistical methods for their applications. However, simulation studies presented by methodological researchers as part of methodological papers typically do not fulfill this role. We claim that efforts should be made by methodological researchers towards translating the simulation studies they conduct into statistical practice. As demonstrated through our two use cases, this can be achieved in various ways, such as through user-friendly Shiny applications that allow applied statisticians to replicate the simulations in their own settings without any coding, or—requiring less effort from the methodologists—by providing code that is structured in a modular way that makes it easy for applied statisticians to input their own parameters and rerun the simulations.
Regardless of the specific form its implementation takes, we believe that when methodological researchers consciously incorporate the concept of translational simulation into their work, it can play a key role in bridging the gap between methodological research and statistical practice.

\section*{Declarations}
\subsection*{Ethics approval and consent to participate}
Not applicable. 

\subsection*{Consent for publication}
Not applicable.

\subsection*{Availability of data and materials}
The R codes are available as supplementary materials. The shiny app is publicly available and the link is provided in the manuscript.

\subsection*{Competing interests}
The authors declare that they have no competing interests.

\subsection*{Funding}
This project was funded by DFG grants 3139/7-2 and 3139/9-1 to ALB.

\subsection*{Authors' contributions}
ALB developed the concept and drafted the manuscript. PC developed the shiny app and co-drafted the manuscript. LH contributed to the implementation of the examples. VG provided the first clinical example. EH contributed to the manuscript. All authors read and approved the manuscript.

\subsection*{Acknowledgements}
We thank Tim Morris, Michal Abrahamowicz and James Pustejovsky for helpful comments.

\end{document}